\documentclass{icrc29}

\usepackage{graphicx,amssymb,amsmath,times}
\begin{document}

\title[Analysis of shower size as estimator of extensive air
shower energy]{Analysis of shower size as estimator of extensive air
shower energy}

\author[Vitor de Souza, Jeferson A. Ortiz, Gustavo Medina-Tanco and
 Federico Sanchez] {Vitor de Souza$^a$, Jeferson A. Ortiz$^a$, Gustavo
  Medina-Tanco$^a$ and  Federico Sanchez$^b$ \\ 
{\it [a] Instituto de Astronomia, Geof\'{\i}sica e Ci\^encias Atmosf\'ericas, Universidade de S\~ao Paulo, Brasil} \\
{\it [b] Dipartimento di Fisica, Universit\'a degli Studi di Milano e I.N.F.N.}
} 

\presenter{Presenter: V. de Souza (vitor@astro.iag.usp.br)}

\maketitle
    
\begin{abstract}

The fluorescence technique has been successfully used to detect
ultrahigh energy cosmic rays by indirect measurements. The underlying
idea is that the number of charged particles  in the atmospheric
shower, i.e, its longitudinal profile, can be extracted from the
amount  of emitted nitrogen fluorescence light.  However the influence
of shower fluctuations  and the very possible presence of different
nuclear species in the primary cosmic ray  spectrum make the estimate
of the shower energy from the fluorescence data analysis a difficult
task. We investigate the potential of shower size at maximum depth as
estimator of shower  energy. The detection of the fluorescence light
is simulated in detail and the reconstruction biases  are carefully
analyzed. We extend our calculations to both HiRes and EUSO
experiments.  This kind of approach is of particular interest for
showers that are not fully contained  inside the field of view of the
detector.
\end{abstract}

\section{Introduction}

The total amount of emitted fluorescence light in a shower is a very good 
approximation to the total number of charged particles $N$($X$), where $X$
is the atmospheric depth. In this sense the number of particles at shower 
maximum can serve as an estimator of the shower energy. The total energy that 
goes into electromagnetic charged particles is obtained by integration of 
the shower longitudinal profile
\begin{equation}
E_{\rm em}=\alpha \int_0^{\infty} N(X)dX
\label{eq:Eem}
\end{equation}
where $\alpha$ is the average ionization loss rate and the integral on the right-hand
side represents the total track length of all charged particles in the shower projected
onto the shower axis. 

As an alternative proposal \cite{dawson02} the electromagnetic energy can also be 
calculated by using the fluorescence light intensity and the fluorescence efficiency,
without the obligation to reconstruct the number of particles as a function of the 
atmospheric depth. Such approach is taken as a very precise measurement of the primary 
shower energy because it is supposed to be weakly dependent of the simulation models 
and the primary particle type. However, when details of the shower development are 
taken into account the calorimetric measurement can lead to high systematic uncertainties.
A not less important concern is that the fluorescence efficiency as a function of air 
pressure, density and humidity is only known up to a certain extent. According the approach given 
by equation \ref{eq:Eem}, the average ionization loss rate is used in the air shower reconstruction 
and hence the energy spectrum of the electron in the shower must be known via Monte Carlo simulation.

Although the electrons and positrons constitute the majority of the charged
particles in a shower and contribute most to the fluorescence light, 
an important fraction of the shower energy is carried by particles which are
invisible to fluorescence telescopes. Such ``missing energy'' is estimated 
using Monte Carlo air shower simulation and contributes to the uncertainties
involved in this method, being sensitive to the primary composition.

Theoretical works have shown the existence of a clear relation between 
the primary energy and the maximum number of particles in the shower. 
Recently, Alvarez-Mu\~niz et al. \cite{Alvarez04} have studied the 
$N_{\mathrm {max}}$ shower quantity as an estimator of the primary shower 
energy, confirming the efficiency of this technique.
However, telescopes particularities and reconstruction procedures must be 
considered due to the introduction of biases and fluctuations in the 
calculation of $N_{\mathrm {max}}$. 

\begin{figure}
\begin{minipage}[t]{7.5cm}
\includegraphics*[angle=-90,width=1.0\textwidth]{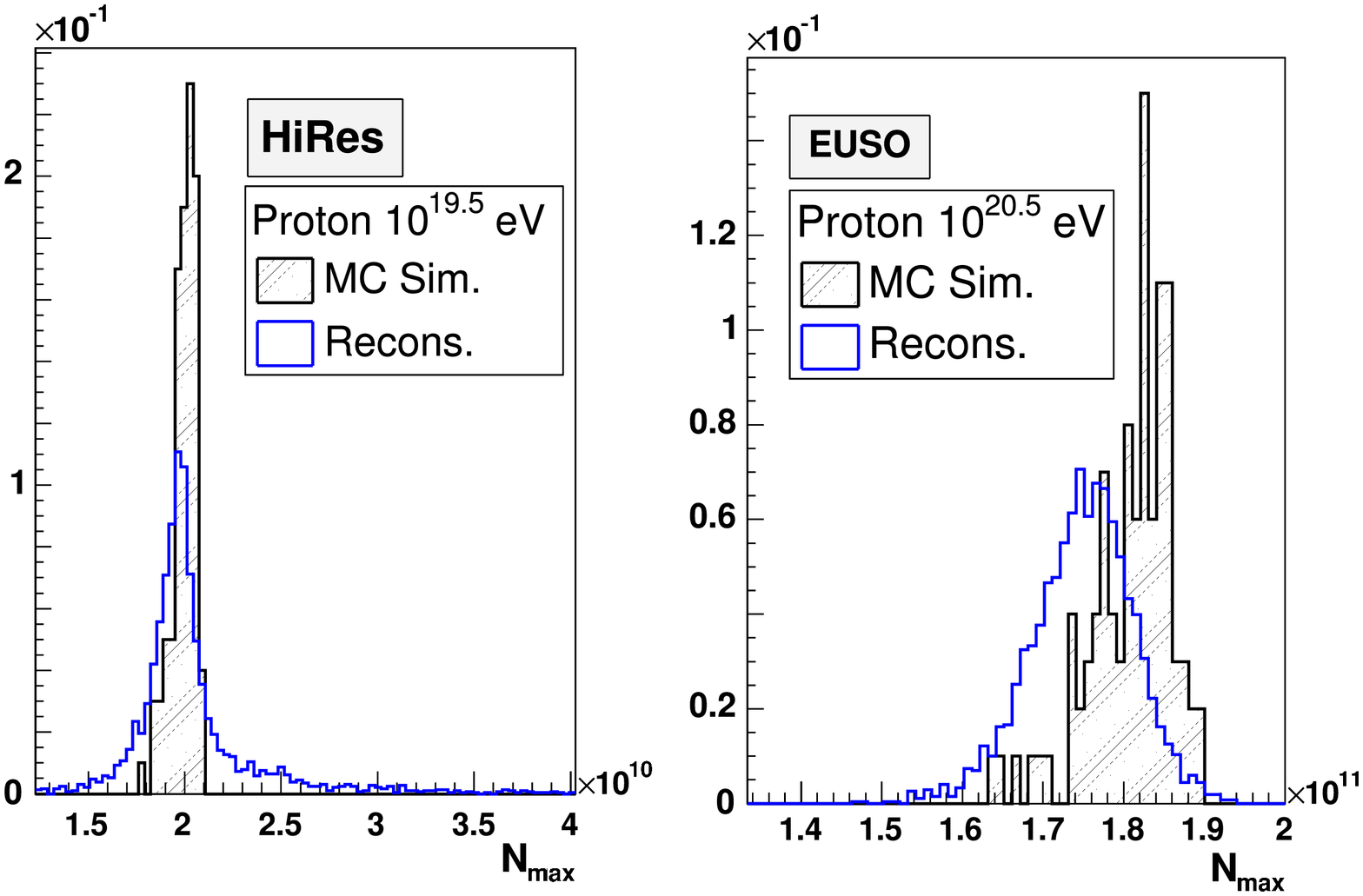}
\vspace{-0.5cm}
\caption{\label{fig:nmax} Normalized distribution of the 
  simulated and reconstructed $N_{\mathrm {max}}$.}
\end{minipage}
\hfill
\begin{minipage}[t]{7.5cm}
\includegraphics*[angle=-90,width=1.0\textwidth]{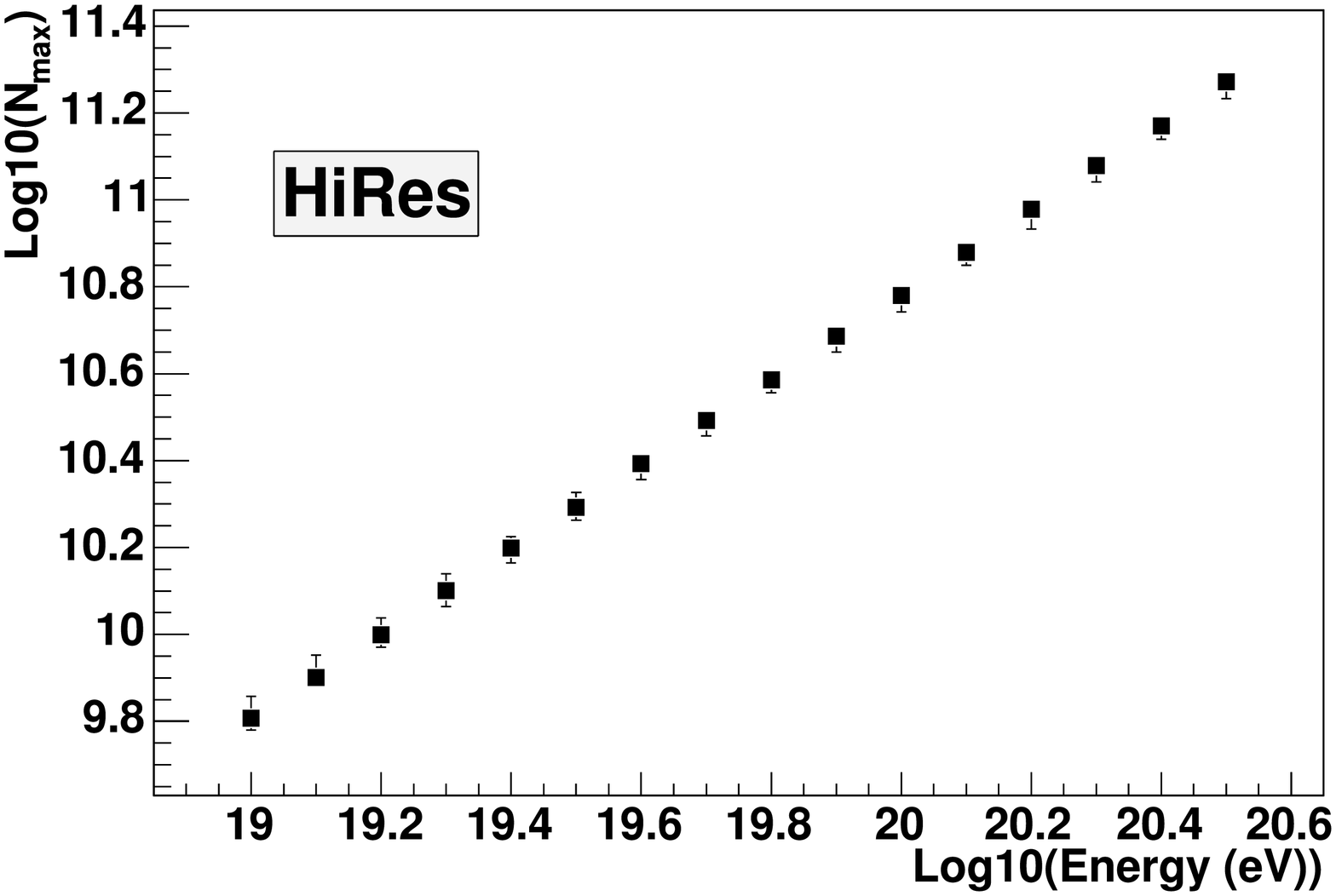}
\vspace{-0.5cm}
\caption{\label{fig:en:nmax:hi} Relation between the reconstructed
  $N_{\mathrm {max}}$ and the simulated energy for the HiRes telescope.}
\end{minipage}
\hfill
\end{figure}

We explore the possibility of estimating the primary shower energy using the $N_{\mathrm{max}}$ 
shower quantity for the HiRes \cite{bib:hires:espectro} and EUSO \cite{Euso} fluorescence experiments, i.e.,
ground and space based experiments, respectively. The telescopes particularities 
and the reconstruction procedures are included in our analysis, predicting more 
realistic results.

\section{Event analysis}

The importance of the fluorescence technique in measuring air showers has been
demonstrated along the years by the Fly's Eye and HiRes collaboration and now is
also being successfully used by the Pierre Auger Observatory \cite{Auger}, all 
ground based experiments. At the same time, projects are under development and
intend to use the fluorescence approach from space observatories. In such projects, 
fluorescence telescopes would be installed at the International Space Station or in
satellites which would increase by at least a factor of 10 the aperture reached by 
the current ground based telescopes.

In this work we study the possibility of using the $N_{\mathrm{max}}$ parameter as 
an estimator of the primary energy, employing HiRes and EUSO telescopes as case studies
to test its quality in two different setups: ground and space based experiments.

The HiRes telescope specifications were considered and simulated 
in complete accordance with \cite{bib:hires:espectro}. 
The program explained in reference \cite{bib:nosso:espectro:icrc} 
has been used again to obtain the comparisons between our simulations 
and the HiRes data and simulations. Such program scheme was adapted 
to mimic in details the EUSO telescope according to the technical 
configuration given in \cite{Euso}. 

The longitudinal air shower profiles were generated by CORSIKA \cite{bib:corsika}
and CONEX \cite{bib:conex} simulators. From the obtained longitudinal particle profiles 
the number of fluorescence photons can be calculated and propagated 
to the telescopes, being in agreement to the general procedure specified in
\cite{bib:fdsim:luis}. 

Once the simulation of the shower and the telescopes has been done, the shower
longitudinal profile is reconstructed in the standard procedure \cite{bib:hires:espectro}. 
The number of fluorescence photons measured as a function of time is 
converted to the number of particles as a function of depth and then a
Gaisser-Hillas profile is fitted. One of the parameters fitted in the
Gaisser-Hillas function was $N_{\mathrm {max}}$. Only showers which survived the
HiRes-II cuts as published in \cite{bib:hires:espectro} were used in the following 
analysis.

The EUSO collaboration has not defined quality cuts yet and therefore we have imposed 
very loose ones requiring total path length greater than $0.6^\circ$ and greater 
than 200~g/cm$^2$ and $X_{\mathrm {max}}$ in the field of view of the telescope.

We have simulated 100 showers with CORSIKA for energies varying from $10^{19.0}$ 
to $10^{20.5}$~eV for the HiRes analysis and with CONEX for energies 
varying from $10^{19.5}$ to $10^{21.5}$~eV for the EUSO. Each shower was used 50
times by drawing a different geometry. For the HiRes calculations we
have randomly distributed the showers with zenith angle smaller than 60$^\circ$
over an area with radius of 50 km. For the EUSO calculations we have drawn core
positions in a circle with radius of 430 km and zenith angle smaller than
$89^\circ$.

Fig.~\ref{fig:nmax} illustrates on the left hand the $N_{\mathrm {max}}$ distribution 
as simulated by the Monte Carlo program and the distribution of the same
reconstructed quantity on the right hand. One can easily verify how the detection and
reconstruction procedures distort the distribution producing a wider
distribution and a shift in the values. Any reconstruction method based on
$N_{\mathrm {max}}$ should take these biases into account. 

\begin{figure}
\begin{minipage}[t]{7.5cm}
\includegraphics*[angle=-90,width=1.0\textwidth]{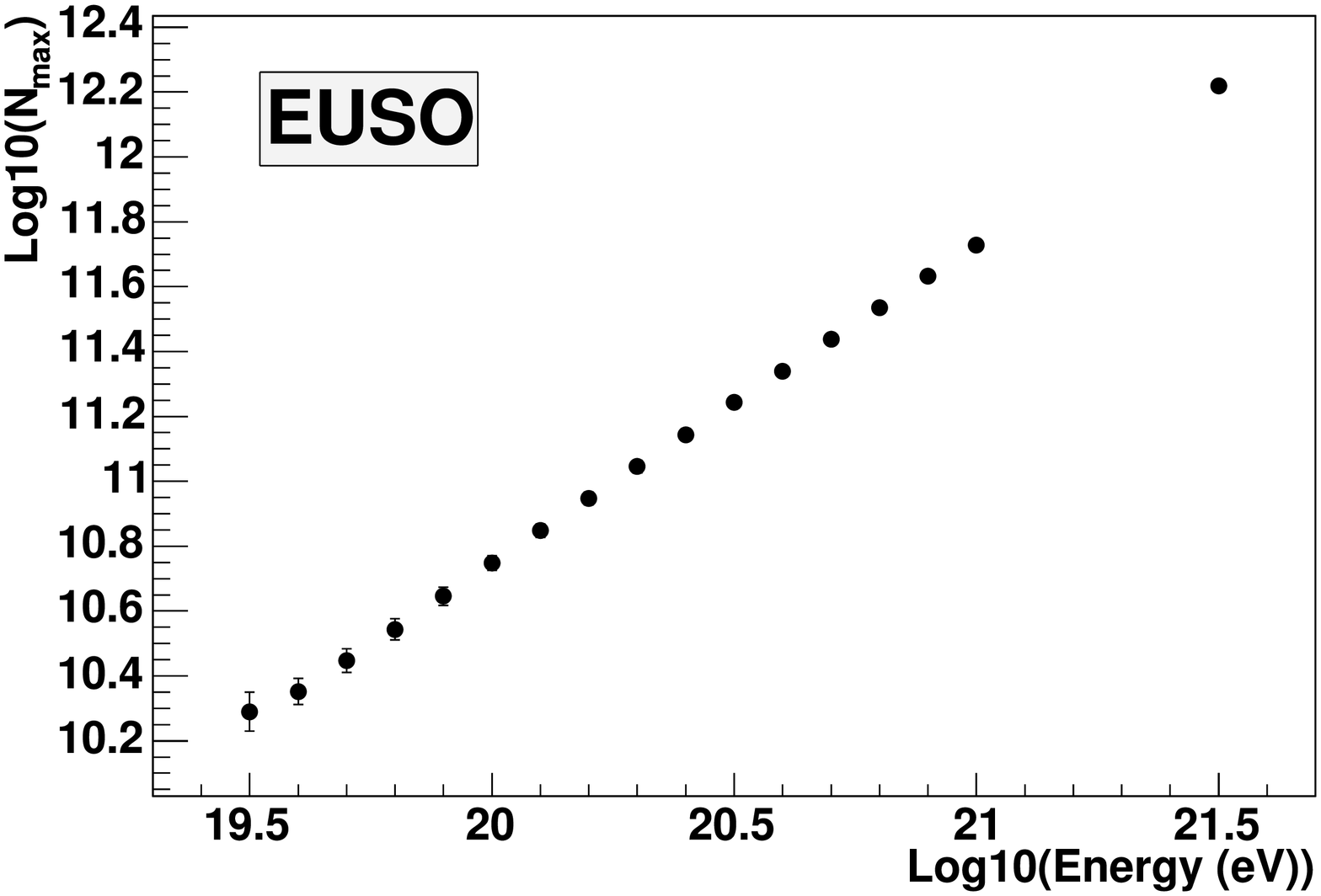}
\vspace{-0.5cm}
\caption{\label{fig:en:nmax:eu} Relation between the reconstructed $N_{\mathrm {max}}$ and the simulated energy for the EUSO telescope.}
\end{minipage}
\hfill
\begin{minipage}[t]{7.5cm}
\includegraphics*[angle=-90,width=1.0\textwidth]{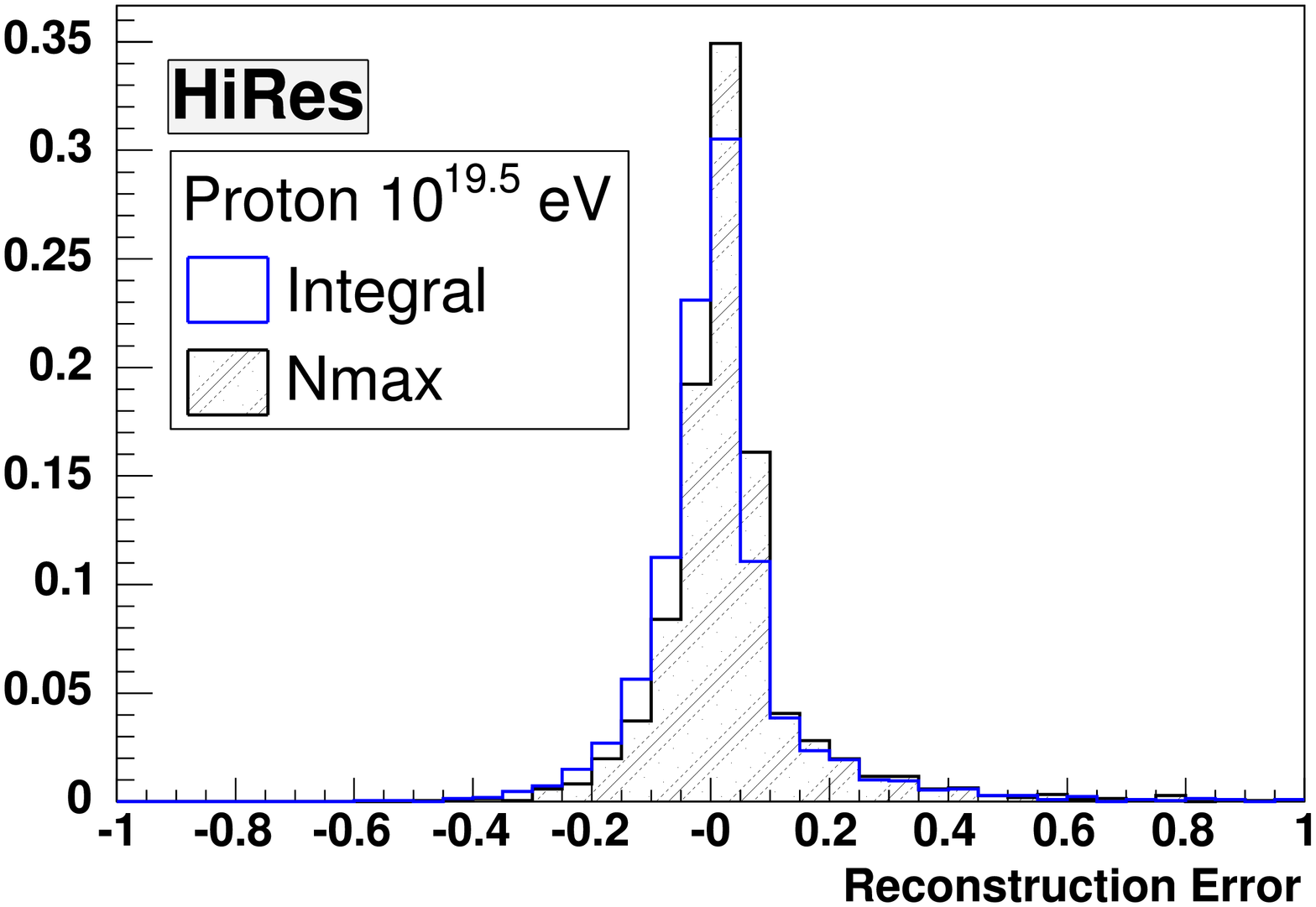}
\vspace{-0.5cm}
\caption{\label{fig:en} Normalized distribution of the error in the energy reconstruction for both methods.}
\end{minipage}
\hfill
\end{figure}

Fig.~\ref{fig:en:nmax:hi} shows the relation between the primary energy simulated 
by the Monte Carlo schemes for the HiRes and EUSO telescopes specifications 
with the average reconstructed $N_{\mathrm {max}}$. The small error bars show 
the one sigma confidence level for the $N_{\mathrm {max}}$ distribution. The 
relation between $N_{\mathrm {max}}$ and energy in Fig.~\ref{fig:en:nmax:hi} can 
be well fitted by a straight line 
\begin{equation}
Energy  = A + B \times N_{\mathrm{max}}
\label{eq:en:nmax}
\end{equation}

The values of \emph{A} and \emph{B} parameters were determined to be $-1.03 \times 10^{18}$ 
and $1.73 \times 10^9$, respectively, for the HiRes experiment and $-5.65 \times 10^{18}$ 
and $1.86 \times 10^9$ for the EUSO telescope. 
  
\section{Results}

Using equations \ref{eq:en:nmax} and \ref{eq:en:nmax} (with the A and B parameters as given above) 
we simulated a second set of showers and reconstructed the energy using both procedures:
a) the standard integral of the Gaisser-Hillas profile (integral) and 
b) the $N_{\mathrm {max}}$ relation (equation~\ref{eq:en:nmax}).

Fig.~\ref{fig:en} illustrates the error distributions related to the energy reconstruction
obtained by using the $N_{\mathrm {max}}$ relation and the integral procedure for the HiRes 
experiment, for proton-induced showers at $10^{19}$~eV.
For the HiRes telescopes the reconstruction error was calculated to be
around 20\% at $10^{19}$~eV reducing to 15\% above $10^{20}$~eV. To this
range of energy the $N_{\mathrm {max}}$ reconstruction showed an
average reconstruction error 1.5\% smaller than the integral procedure. 

\begin{figure}[t]
  \begin{center}
   \includegraphics[angle=-90,width=10cm]{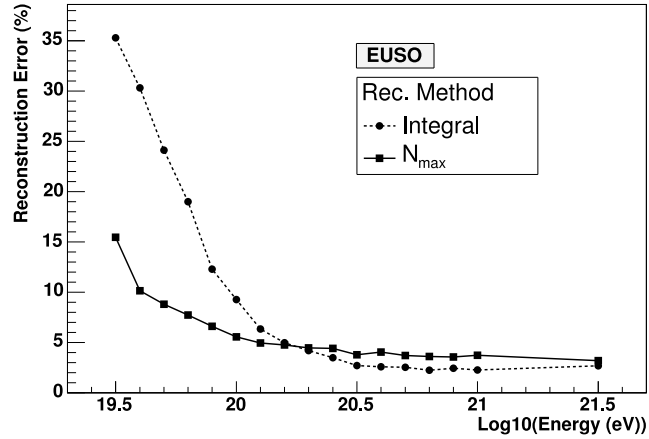}
  \end{center}
  \vspace{-0.5cm}
  \caption{Error in the reconstructed energy for the integral and
  $N_{\mathrm {max}}$  procedures as a function of energy for the EUSO telescope.}
  \label{fig:erro:en}
\end{figure}

Fig.~\ref{fig:erro:en} shows the dependence of the energy reconstruction
as a function of energy for the EUSO telescope. For energies
below $10^{20.3}$ the reconstruction error related to the $N_{\mathrm {max}}$ 
method was smaller. However, for higher energies, the EUSO telescope is 
able to detect the entire development of the shower leading to a good fit 
of the Gaisser-Hillas at all depths and a better efficiency of the energy 
reconstruction.

\vspace{-0.5cm}

\section{Acknowledgments}

This paper was partially supported by the Brazilian Agencies CNPq and
FAPESP. F. Sanchez thanks IAG/USP for its hospitality and INFN/UNIMI
for the funding support. Most of simulations presented here were carried 
on a Cluster Linux TDI, supported by Laborat\'orio de Computa\c c\~ao 
Cient\'{\i}fica Avan\c cada at Universidade de S\~ao Paulo.

\end{document}